\documentclass[twocolumn,aps,prb]{revtex4}
\usepackage{graphicx}
\usepackage{amsfonts}
\usepackage{amsmath}
\usepackage[varg]{txfonts}
\usepackage{bm}
\usepackage{mathrsfs}
\usepackage{subfigure}

\newcommand{\kk}{\ensuremath{{\bf k}}}
\newcommand{\up} {\ensuremath{\uparrow}}
\newcommand{\down} {\ensuremath{\downarrow}}
\graphicspath{{./graphics/}}

\begin{document}

\title{Gutzwiller-projected wave functions for the pseudogap state of underdoped high-temperature superconductors}
\author{Rajdeep Sensarma and Victor Galitski}
\affiliation{Condensed Matter Theory Center and
Joint Quantum Institute, Department of Physics, University of
Maryland, College Park, MD 20742-4111}

\begin{abstract}
Recent experiments strongly suggest that a Fermi surface reconstruction and multiple Fermi pockets are important common features of the underdoped  high-temperature cuprate superconductors. A related theoretical work [Phys. Rev.~B {\bf 79}, 134512 (2009)] has demonstrated that a number of hallmark phenomena observed in the underdoped cuprates appear naturally in the scenario of a paired electron pocket co-existing with unpaired hole pockets. We propose Gutzwiller-projected wave-functions to describe this two-fluid state as well as two competing states in its vicinity. It is argued that a pseudogap state constructed from these wave-functions may be selected by energetics at finite temperatures due to spin fluctuations.
\end{abstract}


\maketitle

Shortly after the discovery of the first high-temperature cuprate
superconductor, Anderson recognized \cite{PWA_Science} strong
electronic correlations as the key to understanding a mechanism for
superconductivity in these compounds and proposed a Gutzwiller
projected BCS wave-function as an approximate ground state. These RVB
ideas have been pursued further in a number of follow-up
works,~\cite{Kotliar_Liu,RVB_Review} which included analytical methods
of renormalized mean-field theory,~\cite{Zhang_Gros} numerical
Monte-Carlo variational analyses,~\cite{Arun,Plain_Vanilla} as well as
effective field theory approaches where non-linear RVB
constraints were implemented via an auxiliary gauge
field.~\cite{Lee_review}  Another key aspect of the cuprate phase
diagram, the presence of an antiferromagnetic (AF) N{\'e}el order at
low dopings, was emphasized later in the framework of the spin-fermion
(SF) model,~\cite{Pines,SFReview} where strong AF fluctuations in the
neighboring metallic normal state were proposed as the pairing glue.


The recent experimental discovery~\cite{QOHTc1,QOHTc2,QOHTc3} of small
Fermi pockets in the underdoped cuprates has initiated a renewed
interest in the physics of these amazing materials and led to new ideas.~\cite{Id1,Id2,Id3,Id4,Id5,Id6,Id7} In particular, the original
breakthrough paper~\cite{QOHTc1} suggested that the Fermi pockets
observed via quantum oscillations might be electron rather than hole
pockets. A later theoretical work~\cite{VGSS} showed that the
``nodal-antinodal dichotomy'' observed in experiment arises naturally
within the picture of a strongly-paired (but uncondensed) electron
pocket in the anti-nodal region and unpaired hole-pockets in the nodal
regions.  The approach of Ref.~[\onlinecite{VGSS}] is to drive a
magnetically ordered AF metal state (with electron and hole
excitations naturally present due to a Brillouin zone folding) into a
critical region via strong AF fluctuations, which were shown to give
way to a strong $s$-wave pairing in the electron pocket and a weaker
$p$-wave pairing in the hole-pockets. The conventional
$d$-wave symmetry is restored via a Brillouin zone unfolding.


Although spin fluctuations are the basic pairing glue in the SF model on 
both the normal Fermi-liquid side~\cite{SFReview} and the AF metal
side,~\cite{VGSS,Moon_Sachdev} they do not induce pair-breaking on the anti-nodal
electrons in the latter model.~\cite{VGSS} The SF model in an AF metal is effectively of Ising
type~\cite{DzeroGorkov} and the $s$-wave electron pairing is the
strongest right in the critical point in a sharp contrast to the
normal Fermi-liquid result, where the pair-breaking dominates at criticality. 
This suggests that it may be energetically favorable to retain the 
AF correlations to a minimum necessary to give
rise to a folding and induce the pairing at criticality. The energy
gain due to the BCS pairing can compensate the presence of otherwise
unwanted AF correlations, such that the charge sector stabilizes a
finite-temperature spin-liquid state.

\begin{figure}[t]
\includegraphics[width=0.4\textwidth]{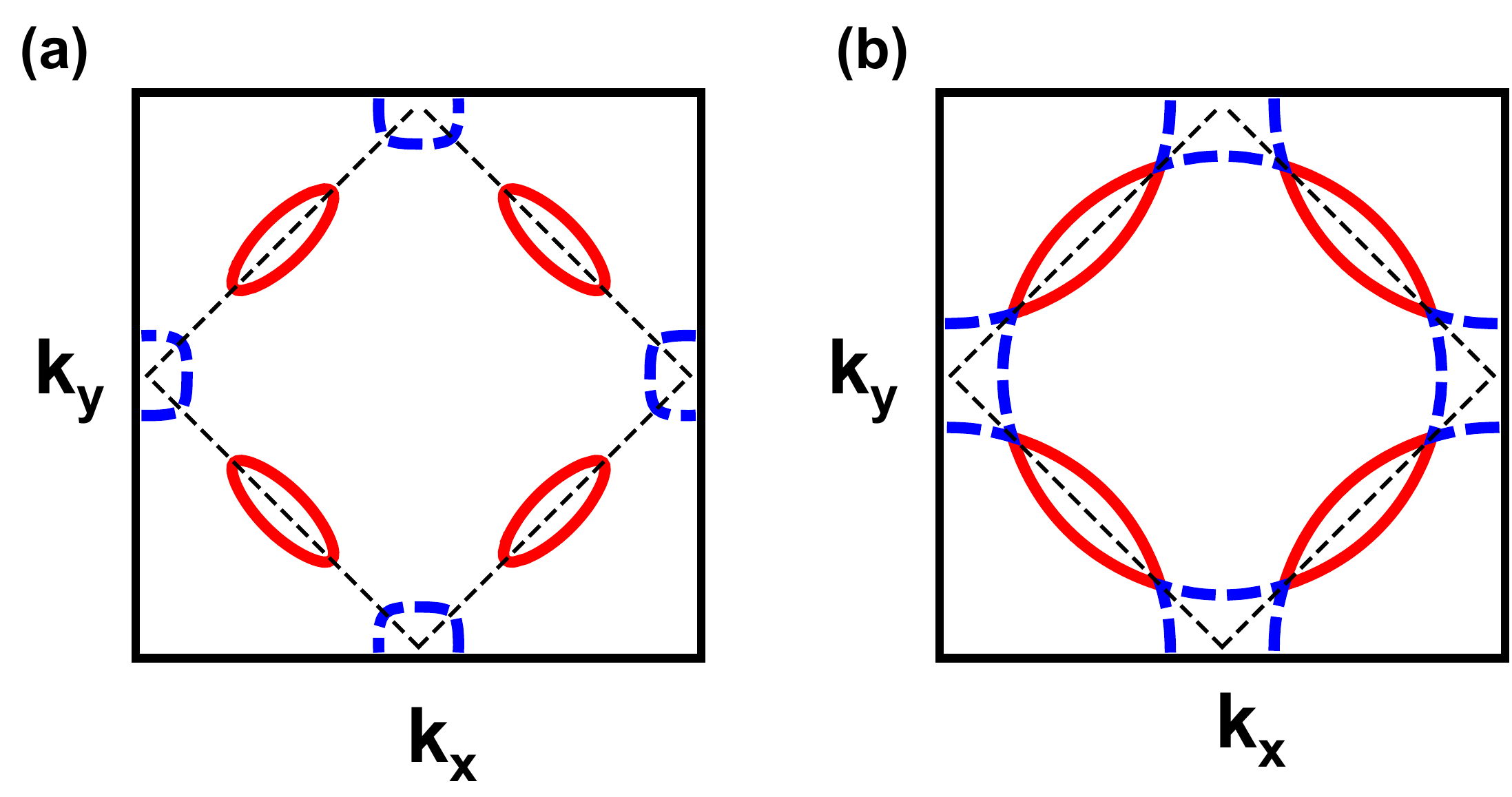}
\caption{(a): Separated nodal hole pockets (solid red line) and antinodal electron pockets (dashed blue line) in a doped antiferromagnet. (b): In the paramagnetic phase the hole and electron pockets touch each other. The thin dashed line marks the magnetic Brillouin zone }
\label{Fig:FS}
\end{figure}
In this paper, we attempt to bring together the SF model and the RVB
theory by proposing a set of Gutzwiller projected wave-functions to  
construct the pseudogap state of Ref.~[\onlinecite{VGSS}].  We
note that our approach below is distinct from other similar
variational wave-function methods used to study competition of
antiferromagnetism and superconductivity,~\cite{Lhuillier,Pathak,Ogata} as
we treat the nodal and antinodal quasiparticles differently. Our
starting point is the $t-t'-J$-model
\begin{equation}
\label{tJ}
\hat{\cal H}_{tJ} = -t \sum\limits_{\left\langle {\bf r} {\bf r}' \right\rangle, \sigma} \hat{c}^\dagger_{{\bf r}\sigma} \hat{c}_{{\bf r}'\sigma}
+t' \sum\limits_{\left\langle\left\langle {\bf r} {\bf r}' \right\rangle\right\rangle, \sigma} \hat{c}^\dagger_{{\bf r}\sigma} \hat{c}_{{\bf r}'\sigma} + {J \over 2} \sum\limits_{\left\langle {\bf r} {\bf r}' \right\rangle} \hat{\bf S}_{\bf r} \cdot \hat{\bf S}_{{\bf r}'},
\end{equation}
where $\hat{c}^\dagger_{{\bf r}\sigma}$ creates an electron of spin $\sigma$ on site ${\bf r}$ and  $\hat{\bf S}_{\bf r} = {1
  \over 2} \sum\limits_{\sigma, \sigma'} \hat{c}^\dagger_{{\bf r}
  \sigma} {\bm \tau}_{\sigma \sigma'} \hat{c}_{{\bf r}\sigma'}$ is the
corresponding spin operator, with ${\bm \tau}_{\sigma
  \sigma'}$ being the Pauli matrices.
Here, $t$ and $t'$ are respectively the nearest and
next-to-nearest neighbor hopping amplitudes, and $J = 4t^2/U$ is the nearest-neighbor AF superexchange coupling, which is related to the on-site Hubbard repulsion, $U$. The $t-t'-J$ model has been widely used to study the effects of strong correlations in the superconducting cuprates. On the other hand, this is also the minimal model which can describe antiferromagnetic correlations due to the superexchange term.

A mean-field Hamiltonian that gives rise to an AF metal state can be constructed from the $t-t'-J$-model (\ref{tJ}) as follows
\begin{equation}
\label{tJMF}
\hat{\cal H}_{AFM} = - \sum\limits_{\left\langle {\bf r} {\bf r}' \right\rangle, \sigma} t_{{\bf r}, {\bf r}'} \hat{c}^\dagger_{{\bf r}\sigma} \hat{c}_{{\bf r}'\sigma}
 - Jz \sum\limits_{{\bf r}}  m {\bf n} \cdot \left[ \left( - 1 \right)^{\zeta({\bf r})} \hat{\bf S}_{{\bf r}} - {m \over 2} {\bf n} \right],
\end{equation}
where $z=4$ is the co-ordination number, $\left( - 1 \right)^{\zeta({\bf r})} = \pm 1$ if ${\bf r} \in S_{A/B}$, with $S_{A/B}$ denoting the A/B-sublattices. In Eq.~(\ref{tJMF}), $m$ is the modulus and the unit vector ${\bf n}$ is the direction of the N{\'e}el order parameter, which we choose to be in the $z$ direction.

Using the Fourier transforms $\hat{a}_{{\bf k},\sigma} = \sum_{{\bf r} \in S_A} \hat{c}_{{\bf r} \sigma} e^{-i {\bf r} \cdot {\bf k}}$ and $\hat{b}_{{\bf k},\sigma} = \sum_{{\bf r} \in S_B} \hat{c}_{{\bf r} \sigma} e^{-i {\bf r} \cdot {\bf k}}$, and the four-component operator $\hat{\vec{\Psi}}^\dagger_{\bf k} = \left( \hat{a}_{{\bf k}, \uparrow}, \hat{b}_{{\bf k}, \uparrow}, \hat{a}_{{\bf k}, \downarrow}, \hat{b}_{{\bf k}, \downarrow} \right)$, the Hamiltonian (\ref{tJMF}) reads
\begin{equation}
\label{tJMF2}
\hat{\cal H}_{AFM} = \sum\limits_{{\bf k}} \hat{\vec{\Psi}}^\dagger_{\bf k}  \Bigl[
\varepsilon({\bf k}) \check{I}_\tau \times \check{I}_\sigma + \varepsilon_1({\bf k}) \check{I}_\tau \times \check{\sigma}_x + \phi \check{\tau}_z \times \check{\sigma}_z \Bigr] \hat{\vec{\Psi}}_{\bf k},
\end{equation}
where $\varepsilon({\bf k}) = -2t \left( \cos{k_x} + \cos{k_y}\right)$
and $\varepsilon_1({\bf k}) = 4t' \cos{k_x} \cos{k_y}$ are the
inter-sublattice and intra-sublattice dispersions, $\check{\bm
  \sigma}$ and $\check{\bm \tau}$ are the Pauli matrices acting
respectively on the sublattice and spin indices, and $\check{I}_\tau$
and $\check{I}_\sigma$ are the $2\times 2$ identity
matrices. Hamiltonian (\ref{tJMF2}) is diagonalized by a Bogoliubov
rotation and yields two eigen-modes, $E_{\pm} ({\bf k}) \equiv E_{e/h}
({\bf k}) = \varepsilon_1({\bf k}) \pm \sqrt{\varepsilon^2({\bf k}) +
  \phi^2}$ with $\phi=Jzm/2$, where ``e'' and ``h'' correspond to
electron- and hole-branches located primarily in the antinodal and
nodal regions of the magnetic Brillouin zone (MBZ) respectively, as
shown in Fig.~\ref{Fig:FS}. The electron and hole operators are given
by
\begin{equation}
\label{eh-ab} 
{\hat{e}^\dagger_{{\bf k} +} \choose \hat{h}^\dagger_{{\bf k} +}}
= \left[ \alpha_{\bf k} \check{I}_\sigma + i \beta_{\bf k} \check{\sigma}_y \right] {\hat{a}^\dagger_{{\bf k} \uparrow} \choose \hat{b}^\dagger_{{\bf k} \uparrow}} \mbox{  and  }
{\hat{e}^\dagger_{{\bf k} -} \choose \hat{h}^\dagger_{{\bf k} -}}
= \left[ \beta_{\bf k} \check{I}_\sigma + i \alpha_{\bf k} \check{\sigma}_y \right] {\hat{a}^\dagger_{{\bf k} \downarrow} \choose \hat{b}^\dagger_{{\bf k} \downarrow}}, 
\end{equation}
where $\alpha^2_{\bf k} = {1 \over 2} \left[ 1 - {\phi / \sqrt{\varepsilon^2({\bf k}) + \phi^2}} \right]$ and $\left|\alpha_{\bf k}\right|^2 + \left|\beta_{\bf k}\right|^2 = 1$. Note that the shapes of the electron and hole  pockets depend on the details of the lattice hoppings via the ratio $t/t'$, and we assume that it is such that both pockets appear at low dopings.~\cite{Altsh_Chubukov,Chubukov_Morr}

We now use operators (\ref{eh-ab}) to express variational wave-functions describing a paired electron pocket and competing metallic and superconducting states. We denote these states by  $\Bigl|m, \Delta_{\rm e}, \Delta_{\rm h} \Bigr\rangle$, where $m$ is the magnetization, and $\Delta_{\rm e/h}$ is the electron/hole variational gap parameter. The simplest such wave-function describes an unpaired AF metal
\begin{widetext}
\begin{equation}
\label{PAFM}
\Bigl|\,{\rm PAFM}\Bigr\rangle \equiv \Bigl|\bar{m}, 0,  0 \Bigr\rangle =\hat{\cal P}_{\rm G} \prod_{{\bf k} \in {\rm MBZ}} \theta \left[ \mu - E_e({\bf k}) \right] \hat{e}_{{\bf k},+}^\dagger\hat{e}_{{\bf k},-}^\dagger \times \prod_{{\bf p} \in {\rm MBZ}} \theta \left[ \mu - E_h({\bf p}) \right] \hat{h}_{{\bf p},+}^\dagger
\hat{h}_{{\bf p},-}^\dagger \Bigl|\,{\rm VAC}\Bigr\rangle,
\end{equation}
where the Gutzwiller projector, $\hat{\cal P}_{\rm G}= \prod_{\bf r} \left( 1 - \hat{n}_{\bf r \up}\hat{n}_{\bf r \down} \right)$, enforces the non-linear no-double-occupancy constraint, $E_{e/h}({\bf k})$ are the spectra given above Eq.~(\ref{eh-ab}), and $\mu$ is the chemical potential. Here and below the overline implies that the corresponding parameter ({\em e.g.}, magnetization, $\bar{m}$, in Eq.~\ref{PAFM}) is calculated self-consistently by minimizing the energy. 

Another possible wave-function, argued to be of relevance to the pseudogap state, is a Gutzwiller-projected product of wave-functions for the paired electron pocket and unpaired hole pockets:
\begin{equation}
\label{PGAP}
\Bigl|\,{\rm PPEP}\Bigr\rangle \equiv \Bigl|m, \Delta_{\rm e},  0 \Bigr\rangle =\hat{\cal P}_{\rm G} \prod_{{\bf k} \in {\rm MBZ}} \left[U_{\bf k}^{(e)} + V_{\bf k}^{(e)}  \hat{e}_{{\bf k},+}^\dagger\hat{e}_{-{\bf k},-}^\dagger \right] \times \prod_{{\bf p} \in {\rm MBZ}}  \, \theta \left[ \mu - E_h({\bf p}) \right] \hat{h}^\dagger_{{\bf p},+} \hat{h}^\dagger_{{\bf p},-} \Bigl|\,{\rm VAC}\Bigr\rangle,
\end{equation}
where $U_{\bf k}^{(e)}$ and $V_{\bf k}^{(e)}$ are the usual Bogoliubov
amplitudes for a pairing in the electron pocket. {\em I.e.}, $\left[
  U_{\bf k}^{(e)}\right]^2 + \left[ V_{\bf k}^{(e)}\right]^2 = 1$ and
$[V_{\bf k}^{(e)}]^2 = {1 \over 2} \left\{ 1 -[ E_e({\bf k})-\mu]/\sqrt{
    [E_e({\bf k})-\mu]^2 + \Delta_e^2({\bf k})} \right\}$. The variational parameter, $\Delta_{\rm e}$, corresponds to the $s$-wave symmetry  of the electron pairing gap in the folded Brillouin zone. But it is an anisotropic $s$-wave in the MBZ, if we identify the antinodal regions connected by a reciprocal lattice vector ${\bf Q}=(\pi,\pi)$. Note that a non-zero $\Delta_{\rm e}$ before projection does not guarantee 
a long-range  order, which is strongly suppressed by the non-linear Gutzwiller projection.~\cite{Arun}

For completeness, we also write down the variational wave-function for an exotic fully-paired state
\begin{equation}
\label{PFPS}
\Bigl|\,{\rm PFPS}\Bigr\rangle \equiv \Bigl|m, \Delta_{\rm e}, \Delta_{\rm h} \Bigr\rangle =\hat{\cal P}_{\rm G} \prod_{{\bf k} \in {\rm MBZ}} \left[U_{\bf k}^{(e)} + V_{\bf k}^{(e)}  \hat{e}_{{\bf k},+}^\dagger\hat{e}_{-{\bf k},-}^\dagger \right] \times  \prod_{{\bf p} \in {\rm MBZ}} \left[U_{\bf p}^{(h)} + V_{\bf p}^{(h)}  \hat{h}_{{\bf p},+}^\dagger\hat{h}_{-{\bf p},-}^\dagger \right] \Bigl|\,{\rm VAC}\Bigr\rangle,
\end{equation}
\end{widetext}
where $U_{\bf k}^{(h)}$ and $V_{\bf k}^{(h)}$ are the Bogoliubov
parameters in the hole pockets. As noted in Ref.~[\onlinecite{VGSS}],
the hole-pairing must have a $p$-wave character in order to reproduce
the known $d$-wave symmetry in the unfolded picture.
 However,only in the
absence of a long-range N{\'e}el AF order and if both the electron and hole gaps are non-zero, 
the state reduces to the more conventional projected $d$-wave BCS state expressed in terms of
the physical electrons $\Bigl|\,{\rm PBCS}\Bigr\rangle =\hat{\cal P}_{\rm
  G} \prod_{{\bf k} \in {\rm BZ}} \left[U_{\bf k} + V_{\bf k}
  \hat{c}_{{\bf k},\uparrow}^\dagger\hat{c}_{-{\bf
      k},\downarrow}^\dagger \right] \Bigl|\,{\rm VAC}\Bigr\rangle$.

We now focus on states (\ref{PAFM}) and (\ref{PGAP}) to calculate their energy  and determine the optimal variational parameters
 in the framework of Gutzwiller renormalized mean-field theory (RMFT).~\cite{Zhang_Gros,RVB_Review,Ogata}  The basic assumption of RMFT is 
that the expectation value of any operator, $\hat{\cal O}$ in the projected state is proportional 
to that in the corresponding unprojected state, {\em i.e.},
$\left\langle\, {\rm P}\psi\, \left| \,\hat{\cal O}\, \right|\,{\rm P}\psi\,\right\rangle \equiv  \left\langle \psi\, \left|\, \hat{\cal P}_{\rm G}\,\hat{\cal O}\, \hat{\cal P}_{\rm G}  \right|\,\psi\right\rangle \approx g_{\cal O} \left\langle \psi\, \left| \,\hat{\cal O} \right|\,\psi\right\rangle$. The proportionality factor, $g_{\cal O}$, is called the Gutzwiller factor.
The first step in building RMFT is to determine the Gutzwiller factors for all  operators in the $tJ$ Hamiltonian (\ref{tJ}). 

We first consider the relation between the on-site densities, $n_{{\bf
    r},\sigma}$, in the projected states and that in the unprojected
states, $n_{{\bf r},\sigma}^{(0)}$.  In an AF-ordered state, we have
$n_{A,\sigma}^{(0)} = n_{B,-\sigma}^{(0)} = {(n_0 / 2)} (1 + \sigma
m)$, where $m$ is the staggered magnetization in the unprojected
states. Note that the unprojected AF metal state is an eigenstate of
the total particle number, hence $\left\langle\, {\rm PAFM}\, \left|
    \,\hat{N}\, \right|\,{\rm PAFM}\,\right\rangle = \left\langle\,
  {\rm AFM}\, \left| \,\hat{N}\, \right|\,{\rm
    AFM}\,\right\rangle$. Therefore, we have in this case $\sum_\sigma
n_{{\bf r},\sigma}^{(0)} = \sum_\sigma n_{{\bf r},\sigma}$, which
leads to $n_{{\bf r},\sigma} = n n_{{\bf r},\sigma}^{(0)} \left[ 1 -
  n_{{\bf r},-\sigma}^{(0)} \right]/ \left[ n - 2 n_{{\bf
      r},\sigma}^{(0)} n_{{\bf r},-\sigma}^{(0)} \right]$, where the
``bare density,'' $n$, is related to the doping level, $x$, simply as
$n = 1-x$.

Consider now the hopping terms $\hat{c}_{{\bf r}, \sigma}^\dagger \hat{c}_{{\bf r}', \sigma}$. In a projected state, this term contributes in a configuration which has hole at site ${\bf r}$ and a spin $\sigma$ at site ${\bf r}'$ and the action of the operator is to reverse this configuration ${\bf r} \leftrightarrow {\bf r}'$. The probabilities for such a process to occur in a projected//unprojected states are $\propto \left[ n_{{\bf r},\sigma} \left( 1 - n_{{\bf r}} // n_{{\bf r},\sigma} \right) n_{{\bf r}',\sigma} \left( 1 - n_{{\bf r}'} //  n_{{\bf r}',\sigma} \right) \right]^{1/2}$. Their ratio determines the Gutzwiller factor for hoppings:\\
$g^{{\bf r}, {\bf r}';\sigma}_t = \left( 1 - n\right) \sqrt{ n_{{\bf r},\sigma} n_{{\bf r}',\sigma}  / \left[
n_{{\bf r},\sigma}^{(0)} \left( 1 - n_{{\bf r},\sigma}^{(0)} \right) n_{{\bf r}',\sigma}^{(0)} \left( 1 - n_{{\bf r}',\sigma}^{(0)} \right)\right]}$.

Using the relation between $n_{{\bf r},\sigma}$ and $n_{{\bf r},\sigma}^{(0)}$ derived above, we obtain the following 
Gutzwiller factors for inter-sublattice hopping $g_t^{AB} \equiv g_1 =2x/\eta_2^{(+)}(x)$ (which is spin-independent), and the spin-dependent inra-sublattice hopping $g_t^{AA;\up} = g_t^{BB;\down}   \equiv g_{2(3)} =2x \eta_1^{(+)}(x) / \left[ \eta_2^{(+)}(x) \eta_1^{(-)}(x)\right]$, and $g_t^{AA;\downarrow} = g_t^{BB;\uparrow}  \equiv g_3 =2x \eta_1^{(-)}(x) / \left[ \eta_2^{(+)}(x) \eta_1^{(+)}(x)\right]$, where we introduced the functions $\eta_k^{(\pm)}(x)
=1 + x \pm m^k (1-x)$ for brevity. The spin-flip renormalization factor can be obtained similarly and reads
$g_s = 4 \left[ \eta_2^{(+)}(x) \right]^{-2}$. Interestingly in the limit of a nearly-perfect N{\'e}el spin polarization, $m \to 1$, at low dopings ($x \ll 1$), we find $g_1 \sim x$, $g_s \sim g_2 \sim 1$, and $g_3 \sim x^2$. This is consistent with the idea by Kane {\em et al.}~\cite{Lee_singlehole} that holes in an AF background are coherent due to intra-sublattice hopping. We note that in evaluating the Gutzwiller factors, we have only taken into account on-site correlations due to projection and neglected non-local correlations of the wavefunctions.~\cite{nonlocal_gutz}

\begin{figure}[t]
\centering
\includegraphics[width=\columnwidth]{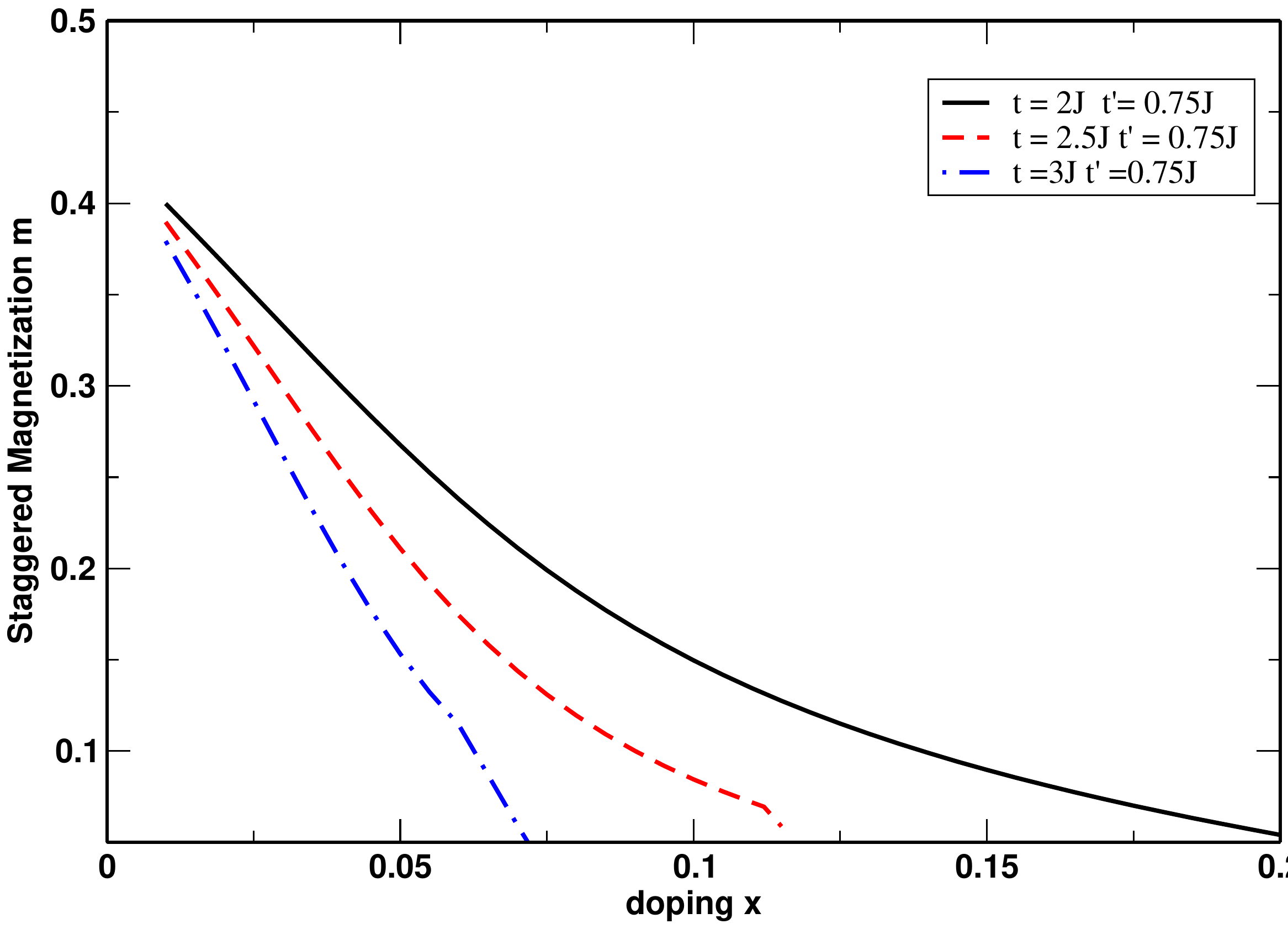}
\vspace*{-0.2in}\caption{ (Color online) N{\'e}el order parameter in the Gutzwiller-projected antiferromagnetic metal~(\protect{\ref{PAFM}}) as a function of doping for three different sets of hopping parameters. In general, a larger $t'$ sustains magnetization over a wider range of
doping levels, as the energy gained from intra-sublattice hopping  favors the AF state. } \vspace*{-0.1in}
\label{Fig1}
\end{figure} 

Using these Gutzwiller factors, we calculate the energy of a generic variational projected state as follows:
\begin{eqnarray}
\label{ENtJ}
\nonumber
&& \left\langle\, {\rm P}\psi\, \left| \,\hat{\cal H}_{\rm tJ}\, \right|\,{\rm P}\psi\,\right\rangle = 
\sum\limits_{{\bf k}\in {\rm MBZ},\sigma;\nu=\pm} S_\nu({\bf k}) \left[ n_e({\bf k},\sigma) + \nu n_h({\bf k},\sigma)\right]\\
\nonumber
&&+\sum\limits_{{\bf k},{\bf k}'\in {\rm MBZ},\sigma} I({\bf k},{\bf k}')   \left[ n_e({\bf k},\sigma) - n_h({\bf k},\sigma)\right] \left[ n_e({\bf k}',\sigma) - n_h({\bf k}',\sigma)\right]\\
&& + \sum\limits_{{\bf k},{\bf k}'\in {\rm MBZ}}  P({\bf k},{\bf k}') \left[ U^{(e)}_{\bf k} V^{(e)}_{\bf k} +U^{(h)}_\kk V^{(h)}_\kk \right]\left[U^{(e)}_{{\bf k}'}V^{(e)}_{{\bf k}'}
+ U^{(h)}_{{\bf k}'}V^{(h)}_{{\bf k}'} \right]\\
\nonumber && + \sum\limits_{{\bf k},{\bf k}'\in {\rm MBZ}} Q({\bf k},{\bf k}') \left[ U^{(e)}_{\bf k} V^{(e)}_{\bf k} - U^{(h)}_\kk V^{(h)}_\kk \right]\left[U^{(e)}_{{\bf k}'}V^{(e)}_{{\bf k}'}
- U^{(h)}_{{\bf k}'}V^{(h)}_{{\bf k}'} \right],
\end{eqnarray}
where $S_+({\bf k}) = \varepsilon_1({\bf k}) \left(g_2 +
  g_3\right)/2$, $S_-({\bf k}) = 2 g_1 \varepsilon({\bf k})
\alpha_{\bf k} \beta_{\bf k} + \varepsilon_1({\bf k}) \left(\beta_{\bf
    k}^2 - \alpha_{\bf k}^2\right) \left(g_2 - g_3\right)/2$, $I({\bf
  k},{\bf k}') = [J g_s/(2t)] \left[(3/2) \varepsilon({\bf k} - {\bf
    k}') \alpha_{\bf k} \beta_{\bf k}\alpha_{{\bf k}'} \beta_{{\bf
      k}'} - t\left(\beta_{\bf k}^2 - \alpha_{\bf
      k}^2\right)\left(\beta_{{\bf k}'}^2 - \alpha_{{\bf k}'}^2
  \right) \right]$, $P({\bf k},{\bf k}') = [3J g_s/(8t)]
\varepsilon({\bf k} - {\bf k}') $, and $Q({\bf k},{\bf k}') = -[J g_s/(8t)]
\varepsilon({\bf k} - {\bf k}')(\beta_\kk^2-\alpha_\kk^2)(\beta_{{\bf k}'}^2-\alpha_{{\bf k}'}^2)$. In Eq.~(\ref{ENtJ}), the
functions $n_e({\bf k},\sigma)$ and $n_h({\bf k},\sigma)$ are the occupation
numbers of the electrons and holes in the corresponding {\em
  unprojected} states. {\em E.g.}, in the AF metal state, they are simply the Fermi-Dirac distributions.

We first focus on the unpaired AF metal state (\ref{PAFM}), where $U_\kk V_\kk = 0$. Minimizing the energy
functional (\ref{ENtJ}), we obtain the staggered magnetization as a function of doping. The magnetization goes down with
doping with the precise slope and the critical doping at which the magnetization vanishes being dependent on the value of $t'/t$, see Fig.~\ref{Fig1}. 

\begin{figure}[t]
\centering
\includegraphics[width=\columnwidth]{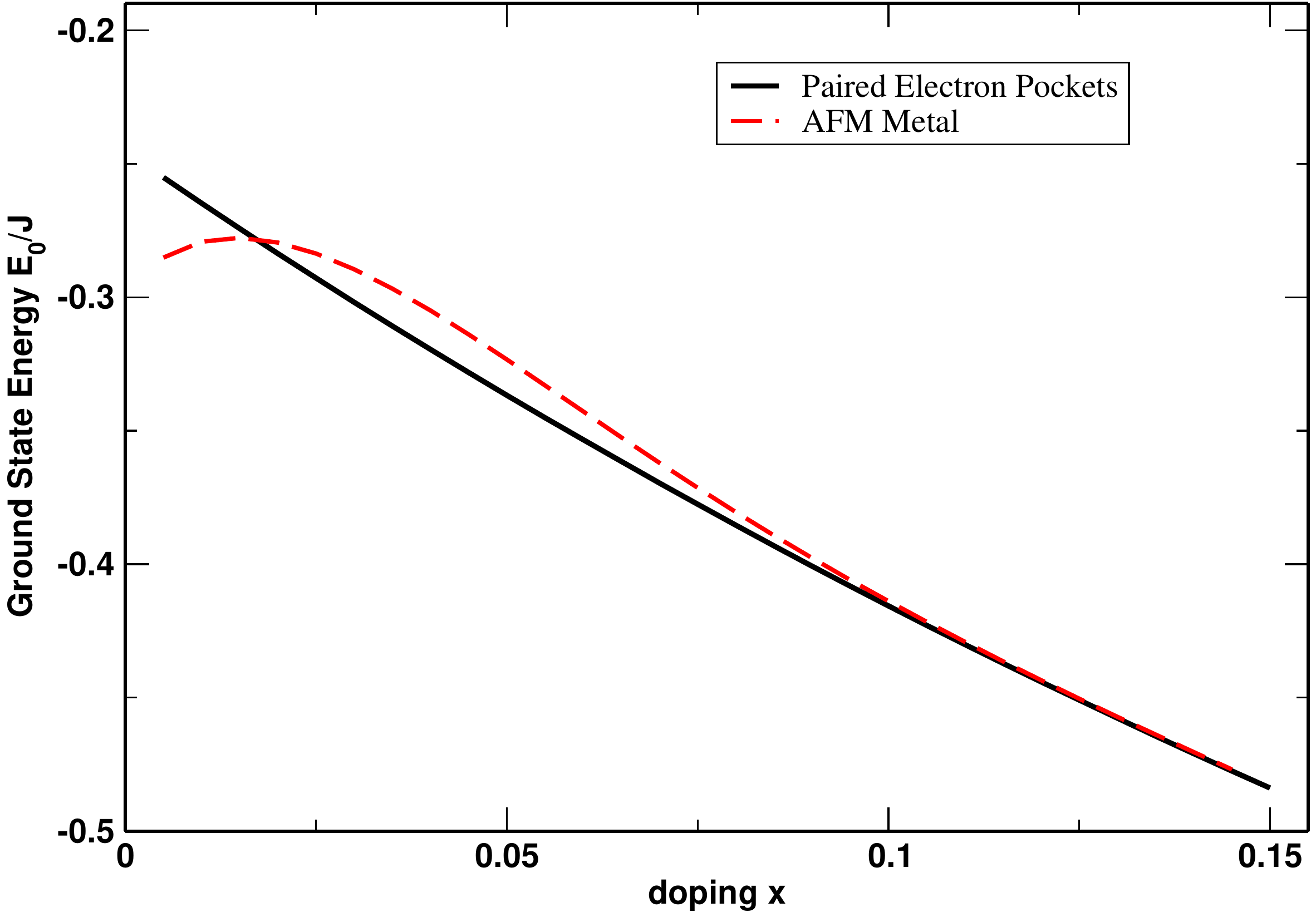}
\caption{(Color online) Displayed are the energy of the paired-electron-pocket state (\ref{PGAP})  and the energy of the projected antiferromagnetic metal (\ref{PAFM}) with optimal magnetization and for the  hopping parameters: $t=2J$ and $t'=-t/4$. Note that at low but finite doping, the two energy curves cross indicating that the state (\ref{PGAP}) becomes energetically preferable to the fully unpaired AF metal (\ref{PAFM}).}
\label{Fig3}
\end{figure}

We next look at  the paired-electron-pocket  wave-function~(\ref{PGAP}). An energetics analysis 
shows the electron gap, $\overline{\Delta}_{\rm e}$,  increasing monotonically with underdoping ({\em c.f.}, Ref.~[\onlinecite{Arun}]) and indicates that small but finite values of $m$ lead to higher energies, hence $\bar{m} \to +0$ is the optimal parameter at the mean-field level. Fig.~\ref{Fig3} compares the $T=0$ energies of the projected AF metal (\ref{PAFM}) and the paired-electron-pocket state (\ref{PGAP}) for a specific reasonable choice of the hopping parameters. Interestingly, the two energies cross, indicating that in the corresponding window of doping levels, a partially paired state is energetically preferable. However, the precise value and even the existence of this window depends strongly on the details of quasiparticle dispersion and hence is non-universal and is strongly susceptible to fluctuation renormalizations. 

Finally, we study the wave-function~(\ref{PFPS}) with both electron and hole pockets
paired in the absence of AF order.  The state reduces to the
projected $d$-wave paired state written in terms of the physical
electron operators and with $(\Delta^e+\Delta^h)$ being
an effective $d$-wave gap. As expected, this fully paired state has
the lowest energy {\em at zero temperature} and yields the projected
$d$-wave superconductor as the ground state.~\cite{Arun,Plain_Vanilla}

However, this does not negate the consideration of
``pseudogap'' wave-function~(\ref{PGAP}) above, since the latter intends to describe
finite-temperature phenomena, which the pseudogap is. Let us invoke
here the arguments of Ref.~[\onlinecite{VGSS}], which showed that spin
fluctuations have a detrimental effect on the $p$-wave pairing of the
holes. Note that the conventional projected wave-function approach
greatly reduces the Hilbert space to that of trial functions, which do
not include excited states. Hence, it can not properly capture 
thermal fluctuations and spatial inhomogeneities. 
However, one can construct a ``variational density matrix'' within this trial Hilbert
subspace
\begin{equation}
\label{DM}
\hat{\rho}_{\rm PGAP} = \int d^3 {\bf m}\, d^2 {\Delta_{\rm e}}\, e^{ - {{\cal F} ({\bf m},\Delta_{\rm e}) \over T}} \, 
\Bigl| {\bf m}, {\Delta_{\rm e}}, 0 \Bigr\rangle \Bigl\langle {\bf m}, {\Delta_{\rm e}}, 0\Bigr|
\end{equation}
to model the impact of (uniform) magnetization and pairing fluctuations. Here, $T$ is temperature and ${\cal F} ({\bf m},\Delta_{\rm e}) = \alpha_m {\bf m}^2  + \beta_m {\bf m}^4 + \alpha_e \left| \Delta_e \right|^2 + \beta_e \left| \Delta_e \right|^4 + \gamma \left| \Delta_e \right|^2 {\bf m}^2$ is a Landau-type functional with $\alpha$'s, $\beta$'s, and $\gamma$ being the variational parameters determined by 
minimizing the energy, ${\cal E}[{\cal F}] = {\rm Tr}\, \left[ \hat{\cal H}_{\rm tJ}\, \hat{\rho} \right] / {\rm Tr}\, \hat{\rho}$, where the trace is over the truncated Hilbert space. Note that in general, a non-zero fluctuating ${\Delta_{\rm h}}$ is also allowed in (\ref{DM}). However, such a state does not reduce to the projected $d$-wave superconductor, and therefore the fluctuating gaps ${\Delta_{\rm e}}$ and ${\Delta_{\rm h}}$ are described by two {\em independent} sets of variational parameters. If AF fluctuations are strong enough, the pseudogap state (\ref{DM}) (with ${\Delta_{\rm h}}$ being strongly suppressed) is selected by energetics. We note here that a proper quantitative analysis of the energetics should include spatial dispersion for fluctuations, restoring a larger Hilbert space and bringing us back to a field theory.~\cite{VGSS} 

In conclusion, we have proposed a set of Gutzwiller projected wave-functions~(\ref{PGAP}) and a related ``variational denisty matrix'' (\ref{DM}), where unpaired nodal holes co-exist with paired antinodal electrons, to describe the pseudogap state of the underdoped high-temperature cuprate superconductors. Finally, we mention that a smoking-gun experiment that could confirm the  pseudo-gap state proposed in Ref.~[\onlinecite{VGSS}] and here would visualize the electron pocket in the underdoped region by suppressing the strong pairing there. Apart from the  transport measurements in ultra-high magnetic fields,~\cite{QOHTc1,QOHTc2,QOHTc3} this could also be achieved by performing ARPES measurements in the non-linear regime of high currents {\em (exceeding the critical current for the electron pocket}) that should destroy a strong s-wave pairing near the antinodes and open up the hidden electron pocket.  

{\em Acknowledgements:} VG is grateful to Professor Anderson for an illuminating discussion. The authors are indebted to Professor Sachdev for reading the manuscript and providing valuable comments. This research was supported by the Department of Energy (VG) and DARPA-QuEST (RS).

\bibliography{RVB}

\end{document}